# Superconducting diode effect under time reversal symmetry

Fengshuo Liu[1,2][†], Yuki M. Itahashi[1]*[†], Shunta Aoki[1], Yu Dong[1], Ziqian Wang[3], Naoki Ogawa[3], Toshiya Ideue[4], and Yoshihiro Iwasa[1,3]*

[1] *Quantum-Phase Electronics Center (QPEC) and Department of Applied Physics, The University of Tokyo, Tokyo 113-8656, Japan*

[2] *Department of Physics, Fudan University, Shanghai 200433, China.*

[3] *RIKEN Center for Emergent Matter Science (CEMS), Wako 351-0198, Japan*

[4] *Institute for Solid State Physics, The University of Tokyo, Kashiwa 277-8581, Japan.*

[†] These authors equally contributed

*Corresponding author: itahashi@ap.t.u-tokyo.ac.jp, iwasa@ap.t.u-tokyo.ac.jp

**In noncentrosymmetric superconductors, superconducting and normal conductions can interchange based on the current flow direction. This effect is termed a superconducting diode effect (SDE), which is a focal point of recent research. The broken inversion and time reversal symmetry is believed to be the requirements of SDE but their intrinsic role has remained elusive. Here, we report strain-controlled SDEs in a layered trigonal superconductor, $PbTaSe_2$. The SDE was found exclusively in a strained device with its absence in an unstrained device, despite that it is allowed in unstrained trigonal structure. Moreover, the zero-field or magnetic field-even (magnetic field-odd) SDE is observed when the strain and current are along armchair (zigzag) direction The results unambiguously demonstrate the intrinsic SDE under time-reversal symmetry and the critical role of strain-induced electric polarization.**

The recently discovered superconducting diode effect (SDE), which is the rectification of supercurrents, has garnered considerable interest from fundamental and practical perspectives (*1-3*). For applications, the SDE should be a fundamental element in superconducting circuits for quantum computing and communication, similar to semiconductor diodes. In terms of fundamental physics, SDE allows the investigation of the relationships among symmetry (*4*), spin-orbit interaction (*5*), vortex dynamics (*6*), and unconventional superconductivity (*7,8*). Consequently, many studies on SDE have been conducted.

A pioneering study related to the V/Nb/Ta tricolor superlattice was reported under an external magnetic field (*9*), and subsequently, studies pertaining to the SDE were mainly performed under an external magnetic field with an odd-parity magnetic field dependence (*10-16*). Meanwhile, SDEs without an external magnetic field have been limited to van der Waals Josephson junctions, twisted interfaces, and magnetization systems. Currently, a fundamental issue regarding the zero-field SDE is raised: the necessity of the breaking of time-reversal symmetry and its origin. If we exclude cases of superconductors proximitized with ferromagnets, which break the time-reversal symmetry (*17,18*), then the remaining examples correspond to Josephson diodes (*19*), twisted trilayer graphene (*20*) and twisted cuprate superconductors (*21*). For Josephson diodes, the asymmetric charging effect owing to electric polarization, even without time-reversal symmetry, has been theoretically discussed (*22-24*). However, the SDE in polar bulk crystals with translational symmetry remains unexplored (*3*).

To discuss the mechanisms underlying the SDE, it is necessary to confirm whether the observed SDE is intrinsic or extrinsic. The former is driven by material properties, that is, the crystal structure and/or unconventional superconducting phases (*9-13,17-18,25-26*), whereas the latter is due to asymmetric device structures (*16,27*) or coupling of the Josephson junction (*15-16,28*). However, these two mechanisms are difficult to distinguish. Thus, the observation of zero-field SDE in more forms, particularly in bulk single-crystal forms without complex device structures, is crucial for identifying the intrinsic mechanism originating from the materials themselves.

In this study, we selected PbTaSe$_2$ with a trigonal space group $P\bar{6}/m2$ (which is an intrinsically noncentrosymmetric superconductor), on which we observed second-order nonlinear longitudinal (measured voltage parallel to current) and transverse (measured voltage perpendicular to current) resistances in both normal and superconducting states at zero-magnetic fields (*29*). A zero-field nonlinear longitudinal response was observed when the current was applied parallel to the armchair direction, whereas transverse nonlinearity was observed when the current was applied parallel to the zigzag direction (*29*). More importantly, these observations of the zero-field nonlinear responses are consistent with crystal symmetry. We attribute this to the scattering of electrons (*30*) instead of the Berry curvature dipole because the trigonal system does not host the Berry curvature dipole (*31*). As exemplified by this example, the crystal symmetry and current direction dependence provide critical information pertaining to the mechanism of nonlinear transport; thus, the study of the zero-field SDE in single-crystal superconductors is crucial.

Another important feature of trigonal symmetry, which is noncentrosymmetric but nonpolar, is that the application of strain induces symmetry reduction in the polar structure. Particularly in two-dimensional van der Waals materials, tensile strain is vital for the engineering of crystal symmetry, and consequently, electronic symmetry, as it enables the realization of novel optoelectronic and electronic properties and functions. Examples include current-induced magnetization in strained monolayer MoS$_2$ (*32*), modulation in the Berry curvature dipole, resultant nonlinear anomalous Hall effect in WSe$_2$ (*33*), and the piezophotovoltaic effect in few-layer 3R-MoS$_2$ (*34*).

Herein, we report a strain-controlled SDE in the trigonal superconductor PbTaSe$_2$. Despite its trigonal symmetry, which allows for the rectification effect, PbTaSe$_2$ in its unstrained state does not exhibit an SDE, whereas an SDE appears only in strained PbTaSe$_2$. This indicates the importance of strain-induced electric polarization for the SDE. Zero-field and magnetic-field ($B$)-even SDE appeared for the current flow along the armchair direction when a tensile strain was applied along the armchair direction. However, when the strain was applied parallel to the zigzag direction, a $B$-odd SDE emerged for the current flow along the zigzag direction. The observed directional dependence is consistent with the crystal symmetry, thus indicating the intrinsic nature of the observed SDE. This result provides solid evidence that the SDE occurs even without broken time-reversal symmetry.

**Basic properties of PbTaSe$_2$**

PbTaSe$_2$ is a superconductor with transition temperature $T_c$ of 3.7–3.8 K (*35,36*) and is known as a topological nodal line semimetal (*37,38*). Figures. 1A and B show schematic crystal structures of PbTaSe$_2$ drawn using VESTA (*39*). It was composed of alternating stacks of 1H-TaSe$_2$ and trigonal Pb layers. Given that each 1H-TaSe$_2$ layer presents a noncentrosymmetric trigonal structure and the stacking direction is the same for all TaSe$_2$ layers, multilayer PbTaSe$_2$ retains this structure (*29,40*). Thus, it offers an ideal platform for investigating bulk SDE.

Single-crystal PbTaSe$_2$ was exfoliated into micro-sized flakes with a thickness of approximately 100 nm. To apply strain to the exfoliated flakes, we employed a previously reported method, as follows (*34*): Two 200-nm-thick SiO$_2$ pads were deposited 10 μm apart in parallel on a SiO$_2$/Si substrate. Subsequently, a PbTaSe$_2$ flake was transferred onto the two pads (Fig. 1C) using a standard transfer technique (see Supplementary Materials Section I for details). By pressing the polycarbonate stamp, the flake was stretched to establish contact with the SiO$_2$/Si substrate (Fig. 1D). As the flake was elongated along the direction perpendicular to the pads, a relatively uniform strain was generated in the area between the two pads where the flake was in contact with the substrate. Au/Ti (240 nm/10 nm) electrodes were deposited on the flake (Fig. 1E), where the current terminals were fabricated on top of the pads, and voltage probes were placed between the pads, where the strain was relatively uniform. In this configuration, the current flowed along the direction of tensile strain. The tensile strain applied via this method was estimated to be 0.2%–0.3% based on an analysis of the second harmonic generation (SHG), as will be explained later (see Supplementary Materials, Section II) (*34*). To fabricate unstrained devices, we deposited Au/Ti (200 nm/10 nm) electrodes onto directly exfoliated flakes on a flat $SiO_2$/Si substrate. Optical images of all unstrained and strained $PbTaSe_2$ devices are shown in Supplementary Materials, Section III. The crystallographic orientation of the exfoliated flakes was first determined from the shape of the crystals (*41*) and was then confirmed via SHG measurements.

First, we describe some basic transport properties of unstrained $PbTaSe_2$ devices. Figure 1D shows the temperature (*T*) dependence of the resistivity at *B* = 0 T in the unstrained device (unstrained sample 2). The superconducting transition temperature, $T_c$, defined as the midpoint of the resistive transition, was 3.6 K. This value is consistent with those reported in previous studies on bulk crystals (*35,36*). Figure 1E shows the voltage–current (*V-I*) characteristics at *T* = 2 K. In *V-I* characteristics, we measured the voltage *V* parallel to the current *I*. The four superconducting transitions in the positive (red line) and negative (blue line) scans were identical, indicating that neither charging (Josephson-junction-like behavior) nor heating occurred in the device. We defined the critical current as the midpoint of the resistive transition, resulting in positive and negative critical currents, $I_{c+}$ and $I_{c-}$, as the device transitioned from the superconducting state to the normal.

## *V-I* characteristics when current flows along zigzag direction

In this section, we focus on the critical currents in the unstrained and strained $PbTaSe_2$ devices along the zigzag direction parallel to the current flow. To confirm the crystal orientation and strain direction, we measured the linear polarization dependence of the SHG signal (Figs. 2A and B) on these samples. The unstrained sample (Fig. 2A) showed a well-developed six-fold rotational pattern, with peak directions corresponding to the armchair directions. In contrast, the strained sample (Fig. 2B) exhibits an anisotropic six-fold pattern. This linear polarization dependence under strain modulation can be described by the following formula:

$$I_{\parallel}(2\omega) = (A \cos 3\phi + B \cos(2\theta + \phi))^2, \tag{1}$$

where $\phi$ is the direction of linear polarization from the armchair direction, $\theta$ is the direction of the uniaxial strain, and $A$ and $B$ are fitting parameters representing the contributions of the trigonal crystal symmetry and uniaxial strain, respectively (*42*). The red lines in Figs. 2A and B show the fitting curves obtained using Eq. (1). The black arrow indicates the direction of the applied strain obtained from the fitting. Based on the optical images of the devices, we confirmed that the current flowed along the zigzag direction in both the unstrained and strained samples and that the strain was applied almost parallel to the current with a misalignment of only 9.7°. Figures 2C and D show the schematic illustrations of the crystal structures of the devices. In the unstrained device (Fig. 2C), three mirror planes exist owing to threefold rotational symmetry. In the strained device (Fig. 2D), only one mirror plane perpendicular to the current remained, which resulted in electric polarization. In this configuration, the mirror plane is perpendicular to the current; therefore, the zero-field diode effect is hindered. We expected SDE to occur under an out-of-plane magnetic field $B$, which breaks the mirror plane. Figures 2E and F (G and H) show the *V-I* curves in unstrained sample 1 (strained sample 1) at $B$ = 0.4 mT and -0.4 mT (0.5 mT and -0.5 mT), respectively. In unstrained sample 1, the difference between the *V-I* curves in the positive and negative current regimes is indiscernible. Meanwhile, in the strained sample 1, we observed a difference between $I_{c+}$ and $I_{c-}$, which is defined as $\Delta I_c = I_{c+} - I_{c-}$. The sign of $\Delta I_c$ depends on the direction of the magnetic field, indicating an odd-order dependence on $B$. This result indicates that the breaking of the $C_3$ symmetry or emergence of in-plane electric polarization leads to the emergence of the SDE.

Additionally, a difference in the sharpness of the superconducting transition between the strained and unstrained devices and an enhancement in the critical current were observed in all our experiments (see Figs. 2 and 3). In unstrained $PbTaSe_2$, vortex string pairs, which are vortex loops penetrating all the layers, is the origin of broad transition without magnetic field (*29*). When the strain is applied, the pinning effect of vortices is enhanced by the induced defects (*43*). We ascribe the relatively sharp transition and larger critical current in the strained device to the enhanced pinning effects of the vortex string pairs. Strained devices with a sharper transition offer an SDE, which enables a transition between superconducting and normal currents at a specific value of |$I$| (see Supplementary Materials, Section VI) over a wider current range, whereas a gradual superconducting transition hinders this phenomenon. It is important to note that the unstrained device, with the current applied along the zigzag direction (unstrained sample 1), also showed weak *B*-odd behavior (see Supplementary Fig. 2D in the Supplementary Materials, Section IV). However, in the unstrained device, the *V-I* curves in the positive and negative current regions overlapped in the middle of the superconducting transition (Supplementary Figs. 4A and B in the Supplementary Materials Section IV), leading to a finite value of $\Delta I_c$; nevertheless, it was much smaller than that of the strained device (Figs. 2G and H). This behavior may be closely related to the second-harmonic signal peak in the superconducting transition, as reported in previous publications (*29,40,44*) (see Supplementary Materials, Section V).

### *V-I* characteristics when current flows along armchair direction

Next, we examined the strained device, applying both current and strain applied in the armchair direction. Figures 3A and B show the angle-dependent SHG intensities in unstrained and strained sample 2, respectively. In contrast to the results shown in Figs. 2A and B, we discovered that the SHG pattern exhibited a peak along the current direction, indicating that the current flowed along the armchair direction. By fitting Eq. (1), we confirmed that the applied strain was almost parallel to the current direction with a misalignment of 1.2°. Figures 3C and D show the schematic of the crystal structures of the unstrained and strained samples, respectively. Figures 3E and F show the *V-I* curves for the unstrained and strained sample 2, respectively, at $B = 0$ mT. Zero-field SDE is observed in the strained device in Fig. 3F. Importantly, $\Delta I_c$ appeared only in the strained device (strained sample 2), even though the zero-field SDE was symmetrically allowed in both the unstrained and strained devices. This is due to the fact that no mirror plane existed perpendicular to the current. To the best of our knowledge, this is the first observation of a zero-field SDE in bulk crystals without a Josephson junction structure or twisted interfaces. Notably, this study marks the initial instance of a zero-field SDE in bulk crystals, distinct from systems employing a Josephson junction structure or twisted interfaces.

**Magnetic field dependence of SDE**

To comprehensively understand the SDE of strained $PbTaSe_2$, we investigated the out-of-plane magnetic field ($B$) dependence of critical currents. The insets in Fig. 4A (B) show a schematic illustration of the crystal structure of $PbTaSe_2$ when strain and current were applied in the zigzag (armchair) direction. In Figs. 4A and B, the *B* dependence of the average critical current, $I_c = (I_{c+} + I_{c-})/2$, is shown. As the magnetic field increased, $I_c$ decreased, indicating the breaking of superconductivity. In Figs. 4C and D, $\Delta I_c$ is plotted as a function of *B*. When *I* was parallel to the zigzag direction (strained sample 1), $\Delta I_c$ exhibited *B*-odd dependence. However, when *I* was parallel to the armchair direction (strained sample 2), a zero-field SDE was observed, as mentioned above, and $\Delta I_c$ showed *B*-even dependence.

This directional dependence of the SDE is consistent with the crystal symmetry as follows: In the device with *I* // zigzag direction, trigonal symmetry breaks mirror plane perpendicular to $\hat{\boldsymbol{y}}$, which is orthogonal to both current (*I* // $\hat{\boldsymbol{x}}$) and magnetic field (*B* // $\hat{\boldsymbol{z}}$). As a result, a non-vanishing triple product $\hat{\boldsymbol{y}} \cdot (\boldsymbol{B} \times \boldsymbol{I})$ leads to a time-reversal asymmetric (*B*-odd) nonreciprocity (*3*). However, SDE observed in the device with *I* // armchair direction does not fulfil such a symmetry constraint specific to the time-reversal asymmetric systems because the mirror plane perpendicular to $\hat{\boldsymbol{x}}$ is vanished. Instead, we can expect zero-field SDE under time-reversal symmetry (also SDE with *B*-even dependence on *B*) when the current and the axis, along which mirror symmetry is broken, are parallel. This result provides unambiguous evidence that the observed SDE is intrinsic as opposed to extrinsic, owing to the unintended asymmetry of the device structure. In addition, we observed a sign change in $\Delta I_c$ with *B*-even dependence, as shown in Fig. 4D. This oscillatory behavior is highly intriguing and may be due to the higher-order effect of the magnetic field; further investigations are warranted to confirm this hypothesis. As *B* increases, the SDE diminishes because of the weakening of the superconducting coherence. Additionally, this magnetic field dependence was reproduced in other devices, and the unstrained devices showed a bare *B*-dependence (see Supplementary Materials, Sections III, IV, and V), further verifying the absence of the SDE in unstrained PbTaSe$_2$.

**Discussion pertaining to the breaking of time-reversal symmetry**

In Josephson junctions without translational symmetry, the breaking of time-reversal symmetry is not necessarily required for the SDE when the nonlinear capacitance is considered in polar junctions (*22,23*). However, in the present case of the strained and polar single-crystal PbTaSe$_2$, translational symmetry is preserved, and the crystal momentum *k* is well-defined. Thus, two electrons with *k* and -*k* are connected via a time-reversal symmetry operation, and a simple symmetry consideration leads us to conclude that the SDE is prohibited under time-reversal symmetry. The present observation of zero-field or *B*-even SDE should be accounted for in two ways: first, the time-reversal symmetry is broken in the superconducting state, and second, zero-field SDE can occur even under time-reversal symmetry.

First, we consider the former effect. Considering the topological nature of PbTaSe$_2$ (*45,46*), it is possible that the time-reversal symmetry could be spontaneously broken in the superconductcting state. However, it is known that this compound exhibits simple *s*-wave behavior in bulk (*35,36,45*). Also, if the SDE is caused by superconductivity with broken time-reversal symmetry, the sign of Δ*I*$_c$ should change between +*B* and -*B* around zero-field as the absolute value of the magnetic field is decreased, similarly to the case of valley-polarized tri-layer graphene (*20*). However, in our measurements of the strained sample 2 (*I* // armchair), Δ*I*$_c$ did not change its sign between the positive and negative magnetic fields, indicating *B*-even behavior. Therefore, spontaneous breaking of time-reversal symmetry is highly unlikely in the present system and thus the present SDE is completely distinguished from previously reported examples of zero-field SDE.

The second interpretation is that the zero-field SDE does not necessarily require a broken time-reversal symmetry. Usually, it is believe that SDE is forbidden under translational symmetry, as discussed above. However, once we introduce electron–electron interactions, SDE is allowed even without the breaking of time-reversal symmetry. A theoretical consideration showed that the electron correlation effects cause a charging effect in the polar direction, which shows an asymmetric dependence on the current direction, and consequently, a zero-field diode effect in the normal state (*47*). In the superconducting state, once we consider the charge fluctuation, which is the cause of the current flow, asymmetric charge accumulation in the polar structure is induced by the electron correlation and results in the SDE even in crystals with translational symmetry. All the experimental results corroborate that this type of SDE is indeed occurring in the present case.

**Conclusions**

In conclusion, we discovered an SDE under time reversal symmetry in symmetry-engineered trigonal superconductor PbTaSe$_2$. Although a simple symmetry consideration allows the SDE to exhibit a trigonal symmetry, our results clearly indicate the absence of an SDE in unstrained PbTaSe$_2$. The SDE was observed only in strained PbTaSe$_2$, where a $B$-odd (zero-field or $B$-even) SDE appeared in the current and strain along the zigzag (armchair) direction. This directional dependence of the SDE is consistent with crystal symmetry, indicating that the observed SDE is an intrinsic effect rather than an extrinsic effect arising from the asymmetry of the device structure. The unprecedented zero-field SDE indicates that broken time-reversal symmetry is not necessarily required; however, the electric polarization and electron correlations play crucial roles in the zero-field SDE. The present results facilitate the further investigation of bulk superconducting diodes and demonstrate the potential applicability of straintronics to superconducting properties.

We are grateful to N. Nagaosa, H. Matsuoka, and Y. Matsuda for fruitful discussions. Y.M.I. was supported by a grant from the Murata Science Foundation. Y. D. was supported by the World-Leading Innovative Graduate Study Program for Materials Research, Information, and Technology (MERIT-WINGS). Z. W. was supported by the RIKEN Incentive Research Project. This study was supported by JSPS KAKENHI (grant no. JP23H00088, JP19H05602, JP22K20352, JP22J22007, and 21K13889), JST FOREST (Grant No. JPMJFR213A) and the A3 Foresight Program.

**Author contributions:** Y.M.I. and Y.I. conceived the study. F.L. and Y.M.I. fabricated the microdevices, performed the transport experiments, and analyzed the data. SHG measurements were performed by S.A. and Y.D., with assistance from Z.W. and N.O. The manuscript is written by Y.M.I., F.L., and Y.I. All authors participated in the discussions.

**Competing financial interests:** The authors declare that they have no competing interests.

**Data and materials availability:** All data required to evaluate the findings of this study are available in the Supplementary Materials.

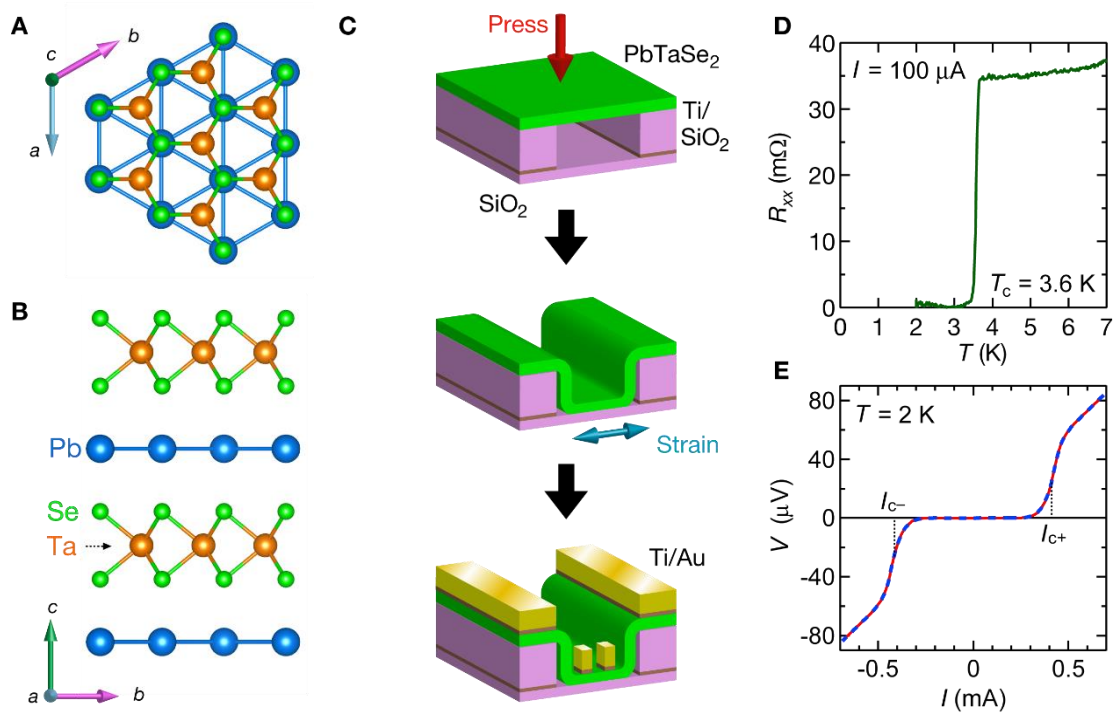

**Fig. 1. Crystal structure, device fabrication process, and basic superconducting properties of unstrained PbTaSe$_2$.** (**A** and **B**) Schematic illustration showing crystal structures of PbTaSe$_2$ from (A) top and (B) side views. Trigonal Pb layers are intercalated in TaSe$_2$ with 1H stacking, resulting in a noncentrosymmetric crystal structure. (**C**) Schematic illustration of device fabrication process. PbTaSe$_2$ flake is pressed on prepatterned SiO$_2$ steps with 200 nm thickness, and strain is applied to the flake via pressing. Au electrodes (240 nm thickness) are evaporated to establish electric contact. (**D**) Temperature dependence of resistance in unstrained device (unstrained sample 2) at $B = 0$ T. Superconducting transition temperature $T_c$ is 3.6 K. (**E**) $V$-$I$ curve for unstrained Sample 2. Red and blue lines indicate positive and negative scans, respectively.

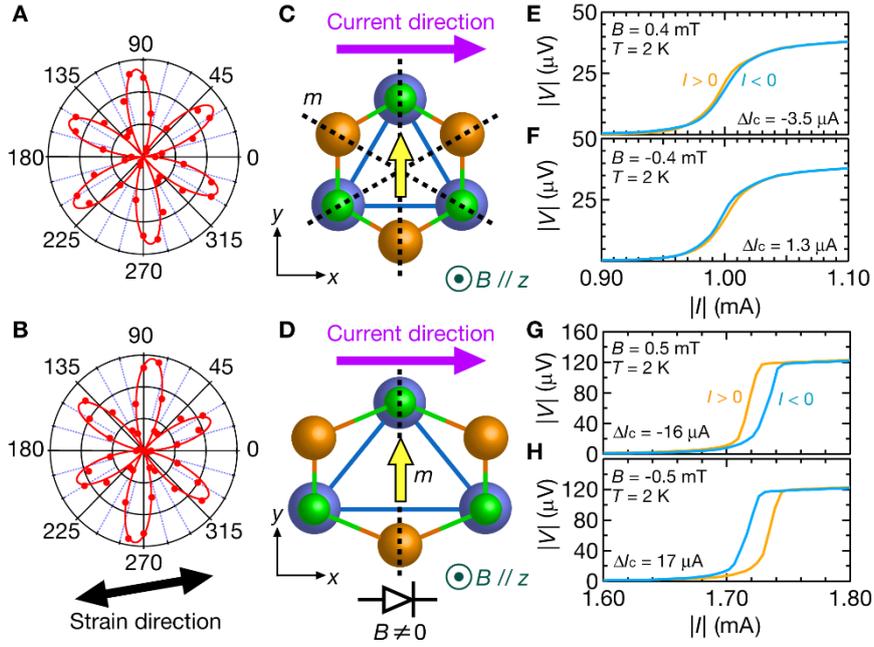

**Fig. 2. *V-I* characteristics of unstrained and strained devices when *I* // zigzag direction.** (**A** and **B**) Laser polarization-resolved SHG intensity patterns of unstrained sample 1 (A) and strained sample 1 (B). Red circles indicate measured SHG intensity and solid red lines indicate fitting curve of SHG intensity pattern obtained using Eq. (1). *B*/*A* = 0.014 and 0.062 in unstrained and strained sample 1, respectively; 0° line is aligned to current direction deduced from optical image. Peak directions in polar plot correspond to armchair direction, thus indicating that current flows along zigzag direction in both samples. Black arrow shows direction of applied tensile strain obtained via fitting. Tensile strain is applied almost parallel to applied current. (**C** and **D**) Schematic illustration of crystal structures of unstrained sample 1 (C) and strained sample 1 (D) assessed based on angular dependence of SHG intensity. In the unstrained device, three mirror planes exists due to three-fold rotational symmetry. In the strained device, only one mirror plane exists perpendicularto the current, thus resulting in polarization. Magnetic field is applied along the out-of-plane direction and yellow arrows indicate the axis along which mirror plane is broken due to trigonal symmetry. Applied current (purple) is set along $\hat{x}$. In this configuration, mirror plane perpendicular to $\hat{y}$ is broken. (**E** to **H**) *V-I* characteristics of unstrained sample 1 (E and F) and strained sample 1 (G and H) at *T* = 2 K. Magnetic fields of ±0.4 mT and ±0.5 mT are applied in unstrained and strained sample 1, respectively. Orange (turquoise) lines indicate positive (*I* > 0) and negative (*I* < 0) current regions in positive and negative scans, respectively. Difference in critical current, or SDE effect, is observed only in strained device (strained sample 1).

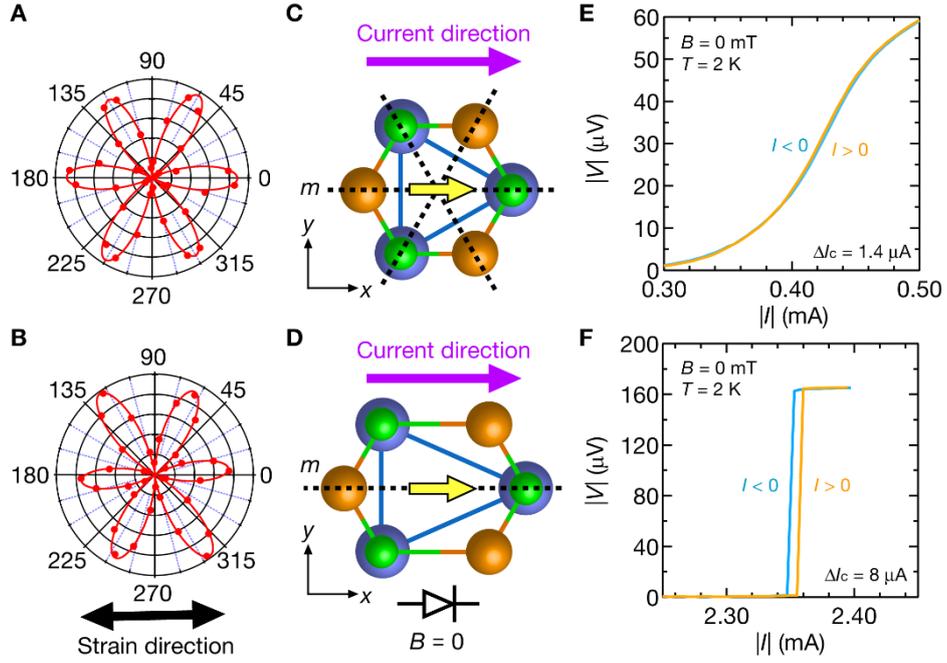

**Fig. 3.** *V-I* characteristics of unstrained and strained devices when *I* // armchair direction. (**A** and **B**) linear polarization dependence of SHG intensity in unstrained sample 2 (A) and strained sample 2 (B). Red circles indicate measured SHG intensity and solid red lines indicate fitting curve of SHG intensity pattern obtained using Eq. (1). $B/A$ = 0.025 and 0.080 in unstrained and strained sample 2, respectively. In both samples, the applied current flows along armchair direction. (**C** and **D**) Schematic illustration of crystal structures of unstrained sample 2 (C) and strained sample 2 (D) assessed from angular dependence of SHG intensity. Yellow arrows indicate the axis along which mirror plane is broken due to trigonal symmetry. Applied current (purple) is set along $\hat{x}$. In this configuration, mirror plane perpendicular to $\hat{x}$ is broken. (**E** and **F**) *V-I* characteristics of unstrained sample 2 (E) and strained sample 2 (F) at $B$ = 0 mT and $T$ = 2 K. Orange (turquoise) lines show positive ($I > 0$) and negative ($I < 0$) current regions in positive and negative scans, respectively. Difference in critical current, or SDE effect, is observed only in strained device (strained sample 2).

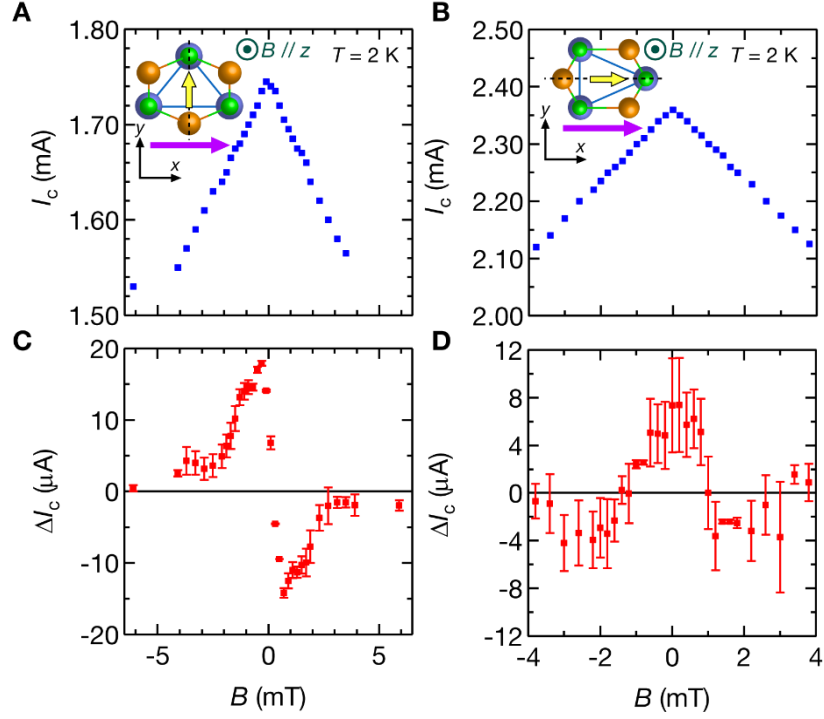

**Fig. 4. Magnetic field dependence of SDE.** (**A** and **B**) Out-of-plane magnetic field ($B$) dependence of averaged critical current $I_c = (I_{c+}+I_{c-})/2$ in strained samples 1 ($I$ // zigzag) (A) and 2 ($I$ // armchair) (B). $I_{c+}$ and $I_{c-}$ are defined as midpoint of resistive transition of positive and negative current regimes, respectively. Insets show schematic illustration of crystal structures. Magnetic field is applied along the out-of-plane direction and yellow arrows indicate the axis along which mirror plane is broken due to trigonal symmetry. Applied current (purple) is set along $\hat{x}$. When the current flows along zigzag (armchair) direction, mirror plane perpendicular to $\hat{y}$ ($\hat{x}$) is broken. (**C** and **D**) $B$ dependence of difference between $I_{c+}$ and $I_{c-}$ ($\Delta I_c = I_{c+} - I_{c-}$) in strained samples 1 (C) and 2 (D). When current flows along zigzag (armchair) direction, $\Delta I_c$ is odd (even) as a function of $B$, which is consistent with the crystal symmetry.

# Supplementary Materials for

# Superconducting diode effect
# under time reversal symmetry

**I. Materials and Methods**

**II. Estimation of the applied strain strength**

**III. Device pictures and summary of devices**

**IV. Magnetic field dependence of critical currents in unstrained $PbTaSe_2$**

**V. Comparison between nonreciprocal transport and superconducting diode effect**

**VI. Half wave rectification**

**VII. Reproducibility of superconducting diode effect**

**I. Materials and Methods**

Commercially purchased single crystals of PbTaSe$_2$ were mechanically exfoliated using Scotch tape. For the unstrained devices, we realised a Hall-bar configuration using Ti/Au (10 nm/240 nm) electrodes on a flake exfoliated on a Si/SiO$_2$ substrate. For the strained devices, we first fabricated two pad structures (40 μm in length and 12 μm in width) arranged 10 μm apart: Ti/SiO$_2$ (10 nm/ 190 nm) (strain samples 1 and 5) and Ti/Au (10 nm/190 nm) (strain samples 2, 3, 4, 6, and 7) on the Si/SiO$_2$ substrate. The exfoliated PbTaSe$_2$ flakes were transferred onto pad structures and pressed onto the substrate using the all-dry-transfer method (*1*). Finally, a Hall-bar configuration was realised using Ti/Au on the transferred flakes. The pattern was fabricated using electron-beam lithography. Pad structures and electrodes were deposited using an evaporator.

Current was applied using a source meter (Keithley 2400) and the voltage was measured using a nanovoltmeter (Keithley 2162) in a Quantum Design Physical Property Measurement System. After measuring the transport properties, we measured the SHG. The intensity and polarization dependence of SHG were measured at room temperature using a custom-developed optical system. A pulsed laser (800 nm wavelength) generated by a femtosecond laser source (1 kHz, 100 fs) was used to irradiate the sample, and the SHG signal at 400 nm was detected using a CCD camera (PIXIS:1024 B). The linear polarization directions of the incident 800 nm and detected 400 nm light were parallel to each other.

## II. Estimation of the applied strain strength

In this section, we estimate the strength of the applied strain. According to reference (*2*), the ratio of two fitting parameters $A$ and $B$ in our uniaxial strain configuration, i.e., only one principal strain, is written as

$$\frac{B}{A} = \frac{(1+\nu)(p_1 - p_2)\varepsilon_{xx}}{(1-\nu)(p_1 + p_2)\varepsilon_{xx} + 2\chi_0}, \tag{S1}$$

where $p_1$ and $p_2$ are the photoelastic parameters, $\varepsilon_{xx}$ denotes the principal strain, $\nu$ is Poisson's ratio, and $\chi_0$ is the nonlinear susceptibility parameter of the unstrained crystal lattice. In the case where the strain effect is weak ($B/A \ll 1$), the first term in the denominator will be much smaller than $\chi_0$ and can be neglected. Thus, eq. S1 leads to

$$\frac{B}{A} \sim \frac{(1+\nu)(p_1 - p_2)\varepsilon_{xx}}{2\chi_0} \equiv \alpha \varepsilon_{xx}, \tag{S2}$$

which is proportional to the strain strength $\varepsilon_{xx}$. Therefore, in our weak strain regime, the $B/A$ represents the relative strain strength.

For the rough estimation of the strain strength, we utilized the parameter of $MoS_2$ (*2*) since this material is well known and belongs to trigonal transition metal dichalcogenide (TMD) materials similar to $PbTaSe_2$. By using $\nu = 0.2941$, $\chi_0 = 4.5$ nmV$^{-1}$, $p_1 = -0.68$ nmV$^{-1}$%$^{-1}$ and $p_2 = -2.35$ nmV$^{-1}$%$^{-1}$, $\alpha$ is estimated as 0.24 %$^{-1}$. Table S1 summarizes the values of $B/A$ and the strain strength $\varepsilon_{xx}$. The strain strength is 0.2-0.3 % for strained devices while 0.06-0.10 % for unstrained devices.

| Sample | Current direction | $B/A$ | $\varepsilon_{xx}$ |
| --- | --- | --- | --- |
| Strained sample 1 | Zigzag | 0.062 | 0.26 % |
| Strained sample 2 | Armchair | 0.080 | 0.33 % |
| Unstrained sample 1 | Zigzag | 0.014 | 0.058 % |
| Unstrained sample 2 | Armchair | 0.025 | 0.10 % |

**Table S1 | Summary of the strength of the strain.** $\varepsilon_{xx}$ is estimated from eq. S2.

**III. Device pictures and summary of devices**

In Fig. S1, we summarize sample pictures used in the main text. Table S2 summarizes the samples, current direction, fabrication process, and magnitude of the superconducting diode effect (SDE). Here, $\eta = \frac{\Delta I_c}{I_c} = 2\frac{I_{c+} - I_{c-}}{I_{c+} + I_{c-}}$ indicates the normalized magnitude of SDE. We confirmed the reproducibility of SDE in 7 strained samples in total. Overall, $\eta$ values in strained devices are larger than ones in unstrained devices, which apparently indicates the absence of SDE in unstrained devices We note that the ratios of our strained devices are not so prominent compared to other systems, e.g., 4.5 % in V/Nb/Ta tricolor lattice (*3*) and 6.2 % in NbSe$_2$-Nb$_3$Br$_8$-NbSe$_2$ Josephson junction (*4*). This might be due to the small polarization in the present strained PbTaSe$_2$.

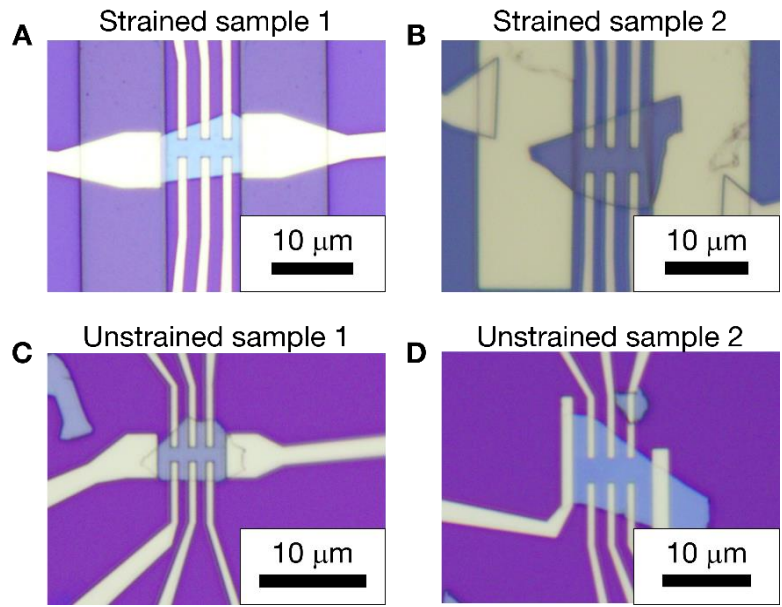

**Fig. S1. Summary of optical images of strained devices.** (**A** and **B**) Optical images of strained samples 1 (A) and 2 (B). (**C** and **D**) Optical images of unstrained samples 1 (C) and 2 (D).

| Sample | Current direction | Step | $\Delta I_c$ | $\eta$ |
| --- | --- | --- | --- | --- |
| Strained sample 1 | Zigzag | SiO$_2$ | 17 mA (0.5 mT) | 1.0 % |
| Strained sample 2 | Armchair | Au | 8 mA (0 mT) | 0.4 % |
| Strained sample 3 | Zigzag | Au | 17 mA (0.2 mT) | 0.9 % |
| Strained sample 4 | Armchair | Au | 8 mA (0 mT) | 0.9 % |
| Strained sample 5 | Zigzag | SiO$_2$ | 12 mA (1.1 mT) | 0.7 % |
| Strained sample 6 | Armchair | Au | 8 mA (0 mT) | 0.4 % |
| Strained sample 7 | Zigzag | Au | 50 mA (0.8 mT) | 2.0 % |
| Unstrained sample 1 | Zigzag | _ | 2.5 mA (0.5 mT) | 0.2 % |
| Unstrained sample 2 | Armchair | _ | 1.4 mA (0 mT) | 0.2 % |
| Unstrained sample 3 | Armchair | _ | 1.7 mA (0 mT) | 0.1 % |

**Table S2 | Summary of all the samples.** Step means the material used as the step structure on the Si/SiO$_2$ substrate. $\eta = \frac{\Delta I_c}{I_c} = 2\frac{I_{c+} - I_{c-}}{I_{c+} + I_{c-}}$ indicates the normalized magnitude of SDE.

**IV. Magnetic field dependence of diode effect in unstrained PbTaSe$_2$**

Since the $\eta$ values of SDE in the present study is relatively small comparing to the those already reported, we have to be careful in drawing conclusions. In this section, we examine the out-of-plane magnetic field dependence in unstrained devices. Figure S2A and C (Figure S3A and C) show average critical current $I_c = (I_{c+} + I_{c-})/2$, where $I_{c+}$ is positive critical current and $I_{c-}$ is negative one, in strained sample 1 (2) and unstrained sample 1 (2), respectively, when *I* // zigzag (*I* // armchair). Figure S2B and D (Figure S3B and D) show the difference between $I_{c+}$ and $I_{c-}$ ($\Delta I_c = I_{c-} - I_{c+}$) in strained sample 1 (2) and unstrained sample 1 (2), respectively, when *I* // zigzag (*I* // armchair). For the case of *I* // zigzag direction, if we only focus on the 50% resistance current, we also observed a *B*-odd behavior with a weaker intensity. However, when we look at the whole *V-I* scan (Fig. S4), you find that the behavior of unstrained devices and strained devices have substantial differences. The two *V-I* scans of the unstrained device (Figs. S4 A and B) have a crossing under a magnetic field, which leads to the sign change of $\Delta V = V(I)-|V(-I)|$ with increasing |*I*|. This sign-change behavior is apparently not characteristic of SDE (Figs. S4 C and D). We deem that this phenomenon is not related to the superconducting transition itself but is the direct result of nonreciprocal transport caused by the dynamics of vortices (*5*) (see section VI in detail). For the case of *I* // armchair direction, the situation is simpler. We observed that the change of magnetic field has little modulation to the difference of critical field in unstrained armchair device, which is apparently different from the behavior shown in strained armchair device. In this case, it is impossible to find a specific current value in which we can switch between a superconducting and normal current depending on the current direction. Taking these results into account, we conclude the absence of SDE in unstrained PbTaSe$_2$.

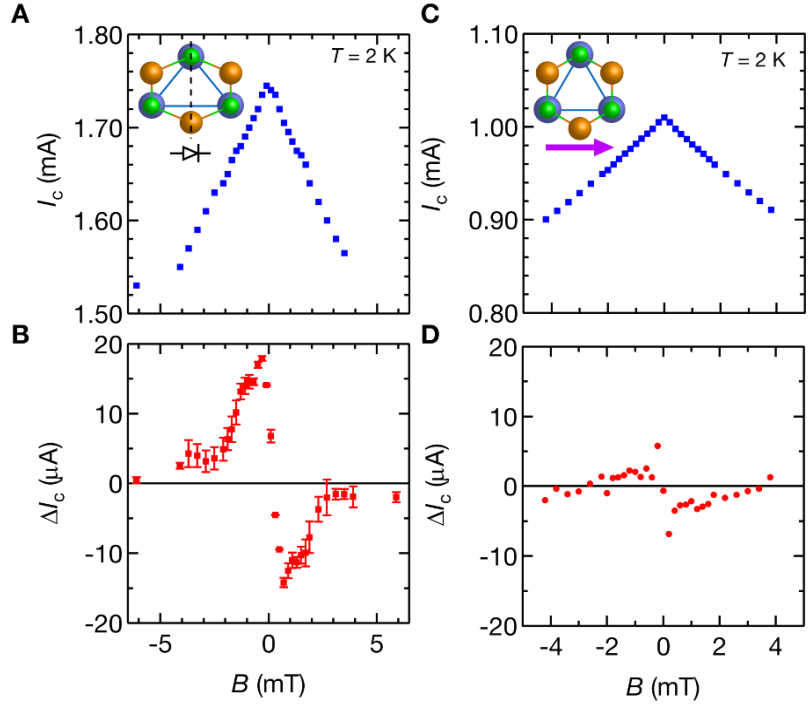

**Fig. S2. Magnetic field dependence of SDE when *I* // zigzag**. (**A** and **B**) Out-of-plane magnetic field (*B*) dependence of average critical current $I_c = (I_{c+} + I_{c-})/2$ (A) and the difference between $I_{c+}$ and $I_{c-}$ ($\Delta I_c = I_{c+} - I_{c+}$) (B), where $I_{c+}$ and $I_{c-}$ are positive and negative critical currents, respectively, in strained sample 1 (*I* // zigzag). (**C** and **D**) *B* dependence of $I_c$ (C) and $\Delta I_c$ (D) in unstrained sample 1 (*I* // zigzag). $I_{c+}$ and $I_{c-}$ are defined as the midpoint of resistive transition of positive and negative current regime, respectively. Insets show the schematic crystal structures.

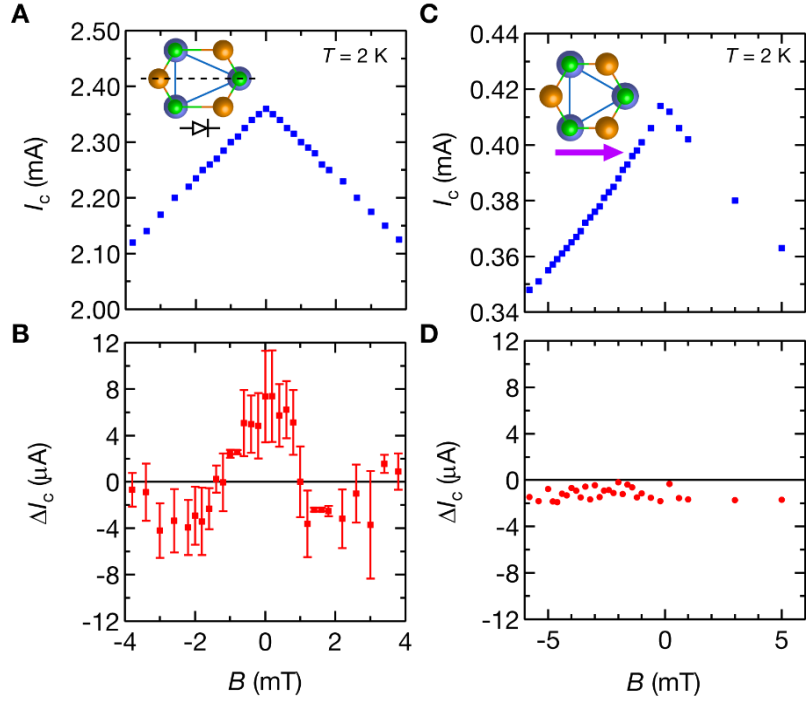

**Fig. S3. Magnetic field dependence of SDE when *I* // armchair**. (**A** and **B**) Out-of-plane magnetic field (*B*) dependence of average critical current $I_c = (I_{c+} + I_{c-})/2$ (A) and the difference between $I_{c+}$ and $I_{c-}$ ($\Delta I_c = I_{c+} - I_{c+}$) (B), where $I_{c+}$ and $I_{c-}$ are positive and negative critical currents, respectively, in strained sample 2 (*I* // armchair). (**C** and **D**) *B* dependence of $I_c$ (C) and $\Delta I_c$ (D) in unstrained sample 2 (*I* // armchair). $I_{c+}$ and $I_{c-}$ are defined as the midpoint of resistive transition of positive and negative current regime, respectively. Insets show the schematic crystal structures.

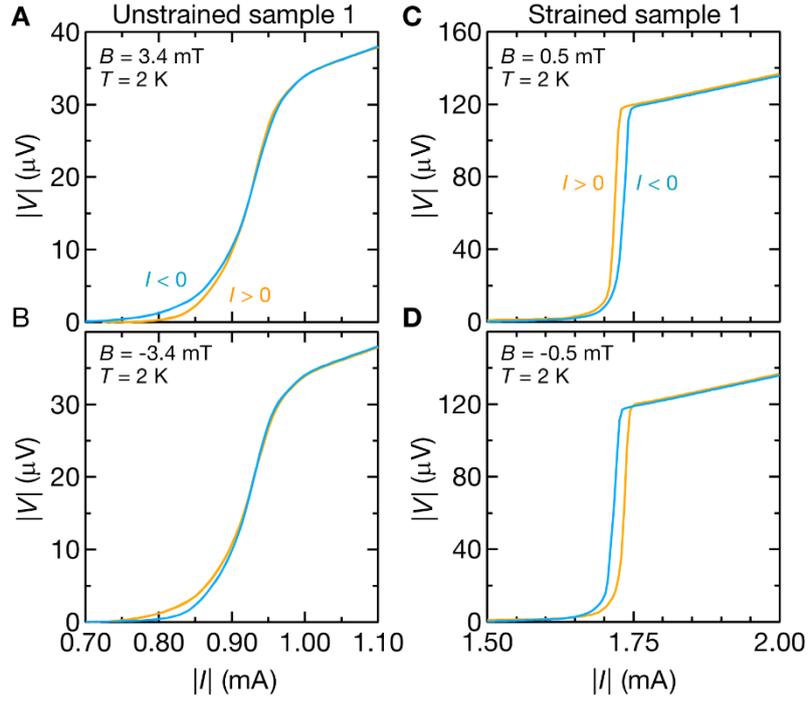

**Fig. S4. Magnification of *V-I* curves when *I* // zigzag**. (**A** and **B**) *V-I* characteristics of unstrained sample 1 (*I* // zigzag) at *B* = 3.4 mT (A) and -3.4 mT (B). (**C** and **D**) *V-I* characteristics of strained sample 1 (*I* // zigzag) at *B* = 0.5 mT (C) and -0.5 mT (D). Orange (turquoise) lines shows positive (*I* > 0) and negative (*I* < 0) current region in positive and negative scans, respectively. In unstrained device, $\Delta I_c$ calculated from 50 % resistance current is negligibly small. On the other hand, in strained device, $\Delta I_c$, or SDE effect is finite.

**V. Comparison between nonreciprocal transport and superconducting diode effect**

We now compare the nonreciprocal transport (*6*) and SDE. In Fig. S5, we show the schematics of the nonreciprocal transport and SDE. The nonreciprocal transport emerges as finite Δ*V* = *V*(*I*)-|*V*(-*I*)| (shown by pink region), which can be detected as second harmonic voltage during the superconducting transition. On the other hand, SDE emerges as difference in critical current Δ$I_c$. In unstrained device, Δ$I_c$ is negligibly small but Δ*V* is finite as mentioned in section III. On the other hand, in strained devices, both Δ$I_c$ and Δ*V* are finite. These explains the presence of nonreciprocal transport (*5*) but the absence of SDE in unstrained PbTaSe$_2$. We also note that the relationship between SDE and Edelstein effect is pointed out (*7*). Edelstein effect in the first order is expected only in polar systems, explaining the presence and absence of SDE in the strained and unstrained PbTaSe$_2$, respectively.

For the comprehensive understanding of the nonreciprocal transport and SDE, we summarized the current direction and magnetic field dependence in Table S3. When we flow the current along the zigzag and armchair directions, the second harmonic nonlinear response under zero magnetic field appears in transverse ($R_{yx}^{2\omega}$, voltage perpendicular to the current is measured) and longitudinal ($R_{xx}^{2\omega}$, voltage parallel to the current is measured) resistance, respectively (*8*). On the other hand, once we apply the out-of-plane magnetic field, *B*-odd second harmonic nonlinear response appears in $R_{xx}^{2\omega}$ and $R_{yx}^{2\omega}$ when the current flows along the zigzag and armchair directions, respectively (*5*). In the present case of SDE, we measured voltage parallel to the current. SDE is not observed in unstrained devices. However, once we apply the strain along current direction, SDE shows *B*-odd and *B*-even (also without *B*) behavior when we flow the current along the zigzag and armchair directions, respectively, similarly to $R_{xx}^{2\omega}$.

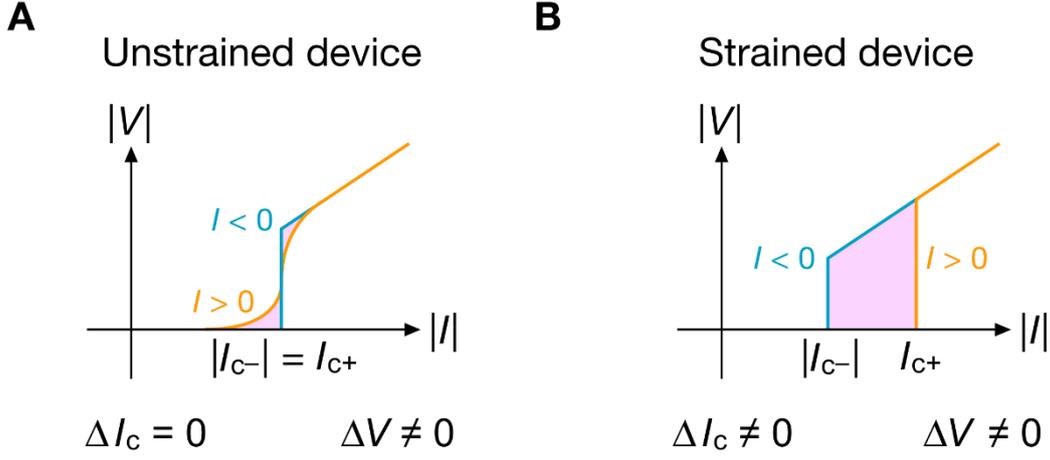

**Fig. S5. Comparison between nonreciprocal transport and SDE.** (**A**) Schematic *V-I* characteristics of unstrained devices. (**B**) Schematic *V-I* characteristics of strained device. Orange (turquoise) line shows positive (*I* > 0) and negative (*I* < 0) current region. Pink region indicates the region with finite $\Delta V = V(I) - |V(-I)|$, which causes the nonreciprocal transport. $\Delta V$ is finite for both cases but $\Delta I_c$ is absent (present) in unstrained (strained) devices.

| Unstrained | *I* // armchair | *I* // zigzag | Reference |
|---|---|---|---|
| $R_{xx}^{2\omega}$ | Without *B* | *B*-odd | Supplementary ref. 5, 8 |
| $R_{yx}^{2\omega}$ | *B*-odd | Without *B* | |
| SDE | — | — | Present work |
| **Strained** | | | |
| SDE | Without *B* (*B*-even) | *B*-odd | Present work |

**Table S3. Summary of the current direction and expected rectification effect for nonreciprocal transport and SDE.** When we flow the current along the zigzag and armchair directions, the second harmonic nonlinear response under zero magnetic field appears in transverse ($R_{yx}^{2\omega}$, voltage perpendicular to the current is measured) and longitudinal ($R_{xx}^{2\omega}$, voltage parallel to the current is measured) resistance, respectively. Under the out-of-plane magnetic field, *B*-odd second harmonic nonlinear response appears in $R_{xx}^{2\omega}$ and $R_{yx}^{2\omega}$ when the current flows along the zigzag and armchair directions, respectively. SDE is not observed in unstrained devices. However, under the application of strain along the current direction, SDE shows *B*-odd and *B*-even (also without *B*) behavior when we flow the current along the zigzag and armchair directions, respectively.

## VI. Half wave rectification

The SDE is further confirmed by the alternately applied dc current. Figure S6A and B show the applied dc current (A) and measured voltage (B) in strained sample 5 ($I$ // zigzag) at $B$ = -2 mT. When the applied current is varied between +1.7 mA and −1.7 mA, the zero-resistance superconducting state and finite-resistance normal state can be switched. This indicates the stability of SDE.

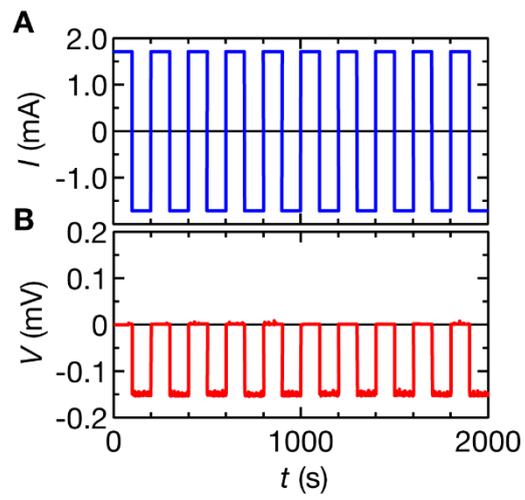

**Fig. S6. Demonstration of half wave rectification.** (**A** and **B**) Switching between the normal and superconducting state demonstrated by alternately applied dc current (A) and measured voltage (B) in strained sample 5 (*I* // zigzag) at *B* = -2 mT. Applied current was 1.7 mA and –1.7 mA.

**VII. Reproducibility of superconducting diode effect**

The reproducibility of SDE is confirmed in other samples. Figures S7A, B, C and D show $B$ dependence of $I_{c+}$ and $I_{c-}$ in strained samples 3 ($I$ // zigzag) (A), 5 ($I$ // zigzag) (B), 4 ($I$ // armchair) (C), and 6 ($I$ // armchair) (D). Insets show the schematic crystal structures. Figures S7E, F, G, and H show $B$ dependence of the difference between $I_{c+}$ and $I_{c-}$ ($\Delta I_c = I_{c+} - I_{c-}$) in strained samples 3 (E), 5 (F), 4 (G), and 6 (H). When we flow the current along the zigzag direction, $\Delta I_c$ is odd as a function of $B$. Although $\Delta I_c$ under the current along armchair direction is not completely even as a function of $B$, (mixture of odd and even parity, due the small deviation of current direction) we unambiguously observed zero-field SDE. This magnetic field dependence, which is consistent with crystal orientation judged from SHG signals, has been observed in multiple samples in addition to the data shown in the main text (Fig. 4). These results further prove the intrinsic SDE coming from the crystal symmetry. Furthermore, the absence and presence of the superconducting diode effect in unstrained and strained samples, respectively, help us rule out the external superconducting diode effect from the Meissner effect and asymmetric edge (*9*) in our samples.

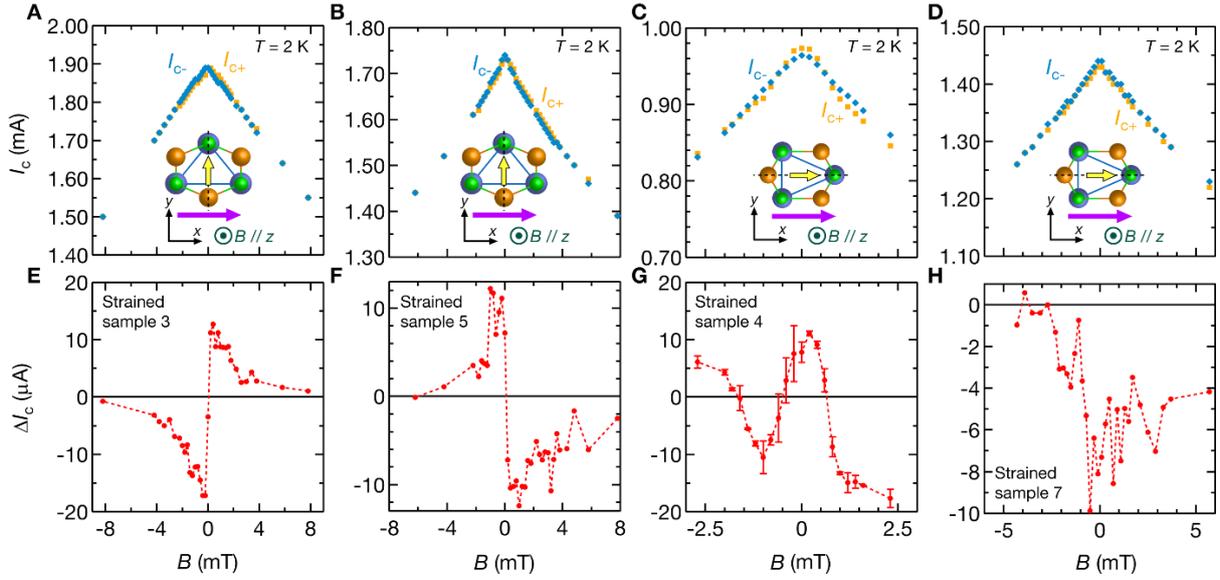

**Fig. S7. Reproducibility of SDE in strained devices**. (**A** and **B**) Out-of-plane magnetic field ($B$) dependence of critical current in positive ($I_{c+}$) and negative ($I_{c-}$) current regime in strained samples 3 (A) and 5 (B) with $I$ // zigzag. (**C** and **D**) $B$ dependence of $I_{c+}$ and $I_{c-}$ in strained samples 4 (C) and 7 (D) with $I$ // armchair. $I_{c+}$ and $I_{c-}$ are defined as the midpoint of resistive transition of positive and negative current regime, respectively. Insets show the schematic crystal structures. (**E** and **F**) $B$ dependence of difference between $I_{c+}$ and $I_{c-}$ ($\Delta I_c = I_{c+} - I_{c-}$) in strained samples 3 (E) and 5 (F) with $I$ // zigzag. (**G** and **H**) $B$ dependence of $\Delta I_c$ in strained samples 4 (G) and 7 (H) with $I$ // armchair. When we flow the current along zigzag (armchair) direction, $\Delta I_c$ is odd (even) as a function of $B$, which is consistent with the crystal symmetry.